\documentclass[12pt]{article}

\usepackage{times}
\usepackage{scicite}

\usepackage[utf8]{inputenc}
\usepackage{comment}
\usepackage{url}
\usepackage{graphicx}
\usepackage{amsmath}
\usepackage{amsthm}
\usepackage{amsfonts}
\usepackage{xr}
\usepackage{amssymb}
\usepackage{lipsum}
\usepackage{floatpag,mwe}
\floatpagestyle{empty}
\usepackage{color}
\usepackage{float}
\usepackage{verbatim}
\usepackage{cases}
\usepackage{longtable}
\usepackage[many]{tcolorbox}
\usepackage{hyperref}

\newtheorem{theorem}{Theorem}

\NewTColorBox{NewBox}{ s O{!htbp} }{%
  floatplacement={#2},
  IfBooleanTF={#1}{float*,width=\textwidth}{float},
  colframe=black,colback=white,
  fonttitle=\fon{phv},
  }

\newenvironment{sciabstract}{%
\begin{quote} \bf}
{\end{quote}}

\title{Repairing dynamic models: a method to obtain identifiable and observable reparameterizations with mechanistic insights} 

\author
{Gemma Massonis,$^{1}$ Julio R. Banga,$^{1}$ Alejandro F. Villaverde$^{1\ast}$\\
\normalsize{$^{1}$BioProcess Engineering Group, IIM-CSIC, Vigo 36208, Galicia, Spain}\\
\normalsize{$^\ast$Correspondence:  afvillaverde@iim.csic.es.}
}

\date{}

\begin{document} 

\baselineskip24pt

\maketitle 

\begin{sciabstract}
Mechanistic dynamic models allow for a quantitative and systematic interpretation of data and the generation of testable hypotheses. However, these models are often over-parameterized, leading to non-identifiability and non-observability, i.e. the impossibility of inferring their parameters and state variables. The lack of structural identifiability and observability (SIO) compromises a model’s ability to make predictions and provide insight. Here we present a methodology, AutoRepar, that corrects SIO deficiencies automatically, yielding reparameterized models that are structurally identifiable and observable. The reparameterization preserves the mechanistic meaning of selected variables, and has the exact same dynamics and input-output mapping as the original model. We implement AutoRepar as an extension of the STRIKE-GOLDD software toolbox for SIO analysis, applying it to several models from the literature to demonstrate its ability to repair their structural deficiencies. AutoRepar increases the applicability of mechanistic models, enabling them to provide reliable information about their parameters and dynamics.
\end{sciabstract}

\section*{Introduction}
Dynamic mathematical models are used for understanding, analyzing, and making quantitative predictions about the behavior of a system over time. They are routinely used in the physical sciences and engineering, and they have also become a fundamental tool in the life sciences\cite{distefano2015dynamic,chen2010classic}, with applications ranging from pharmacology \cite{meibohm1997basic} to medicine \cite{cobelli2019introduction,butner2020mathematical} and biotechnology \cite{almquist2014kinetic,wiechert2011mechanistic}. 
The dynamics of many biological systems or processes of interest can be accurately captured by ordinary differential equations (ODEs) that are generally nonlinear.
In particular, kinetic models are based on first-principles and provide a mechanistic description of the underlying physicochemical processes. Due to their mechanistic nature, they have a significant predictive power \cite{tummler2018discrepancy,alkan2018modeling,arigoni2019mechanistic}, and provide complementary strengths to machine learning approaches \cite{baker2018mechanistic}. However, they usually contain unknown parameters, which must be estimated by optimizing the model fit to experimental measurements of the model output \cite{walter1997identification}. 

This task, known as model calibration\cite{jaqaman2006linking,heinemann2016model,ashyraliyev2008systems}, can only be performed successfully if the optimization problem has a unique solution (i.e., the model is identifiable). Unidentifiable models can lead to false interpretations and misguided interventions \cite{chin2011structural,janzen2016parameter,villaverde2017dynamical,ryser2018identification,schmidt2020recognizing}. 
The impossibility of fulfilling the identifiability condition may be due to an inadequate structure of the model equations (structural unidentifiability) or to insufficiently informative data (practical unidentifiability). The first type of cause must be analyzed theoretically before attempting to calibrate the model in order to characterize the sources of unidentifiability correctly \cite{walter1997identification,distefano2015dynamic}. The second type is dealt with using statistical methods, and it can only be assessed numerically from experimental data. It is important to note that, although checking practical identifiability can be easier than structural identifiability, the latter cannot be deduced from the former \cite{janzen2016parameter}.

Many methods have been proposed to study structural identifiability, starting with the Laplace transformation \cite{bellman1970structural} and including the Taylor series \cite{pohjanpalo1978}, direct tests \cite{denis-vidal:00,walter2004guaranteed}, generating series \cite{walter1982global}, differential algebra \cite{bellu2007daisy,ljung1994global}, power series \cite{pohjanpalo1978system}, observability-based rank tests \cite{sedoglavic2002probabilistic,karlsson2012efficient,villaverde2016structural}, and similarity transformations \cite{evans2002identifiability,yates2009structural}. The analysis of this property can be very complex and in most cases must be carried out computationally. To this end, a number of software tools that implement the above mentioned algorithms have been developed, including ObservabilityTest \cite{sedoglavic2002probabilistic}, DAISY \cite{bellu2007daisy}, GenSSI \cite{Chis11b,ligon2018genssi}, EAR \cite{karlsson2012efficient},  COMBOS\cite{meshkat2014finding}, STRIKE-GOLDD \cite{villaverde2016structural}, and SIAN \cite{hong2018global}.

Structural identifiability (the possibility of uniquely estimating the values of the parameters from the model output) is closely related to another property, observability, which refers to the possibility of inferring the state of the system. If a model is not observable, it cannot be used to predict the time-course of its non-observable states.
We discuss the implications of this scenario in Box 1, where we illustrate the issues with an example from the literature.
The essential point here is that lack of structural identifiability or observability leads to wrong insights and predictions, drastically compromising the usefulness and reliability of the model. 

\begin{NewBox}
\textbf{Box 1. Relevance of structural identifiability and observability}\\
Illustration of how a lack of structural identifiability or observability leads to wrong insights and predictions. The $\beta$IG model shown in diagram (A) represents the glucose-insulin regulation circuit, which maintains plasma glucose concentration within admissible levels \cite{Karin2016DynamicalCompensation}. The glucose uptake ($u$) increases glucose concentration ($G$), which in turn affects the beta cells mass ($\beta$). Beta cells secrete insulin ($I$) at a rate $p$, and insulin decreases glucose concentration at a rate $s_i$. This dynamic system can be modelled with the set of ordinary differential equations shown in panel (B), which contain five unknown parameters ($p,s_i,c,\alpha,\gamma$).
\begin{figure}[H]
\centering
  \includegraphics[width=0.9\linewidth]{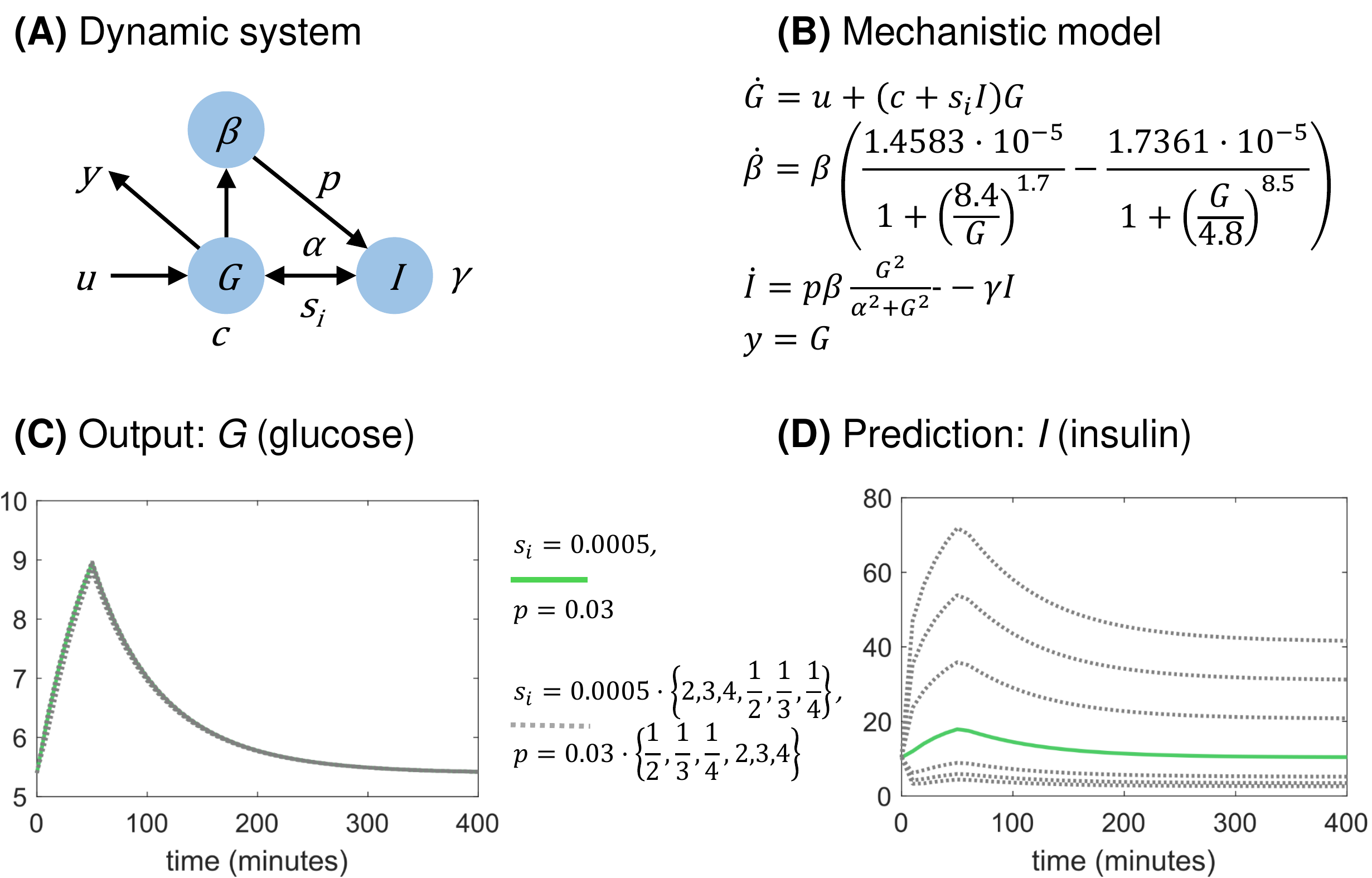}
 \hypertarget{fig:issues}{}
\end{figure}
If we measure glucose concentration over time (i.e. glucose is the model output, $y=G$), two of these parameters $(p, s_i)$ are structurally unidentifiable. In other words, there is an infinite number of pairs of $(p, s_i)$ values that generate the same time-course of glucose. Plot (C) shows that the model simulations of $G$ for several such pairs are identical. Hence, it is not possible to infer the correct values $(s_i=0.0005,p=0.03)$ from glucose measurements. Furthermore, this lack of identifiability is linked to a lack of observability of the unmeasured state variables ($I,\beta$). As a consequence, predictions of the time-courses of ($I,\beta$) made by a model calibrated from experimental data are sure to be wrong. Plot (D) shows the time-courses of the insulin concentration corresponding to the $(p, s_i)$ pairs of panel C, which can differ considerably from the true one (shown in green) even for reasonable values of $(p, s_i)$. The cause of this ambiguity is structural: it is rooted in the existence of symmetries in the model equations. These symmetries are due to over-parameterization: the model contains more parameters that can be estimated from output measurements, and this excessive flexibility hampers its predictive capabilities. 
\end{NewBox}

In this scenario the model user has several options, which are set out in Box 2. 
The ideal solution is to reformulate the model, modifying its dynamic equations in order to remove redundant parameters. In this way we obtain a reparameterized model that is structurally identifiable and observable. Unfortunately, achieving this goal is far from trivial, for two reasons:

\begin{NewBox}
\textbf{Box 2. What can be done with a structurally unidentifiable and/or non-observable model?}
The available options can be classified as follows:
\begin{enumerate}
    \item Continue using the model regardless of its lack of identifiability. This is the easiest yet the least recommendable option, since it may yield modelling artifacts.
    \item Modify the model output, i.e. measure more of their states or functions. An adequate choice of the set of outputs of a model may improve its identifiability \cite{anguelova2012minimal}. However, it is not always possible or practical to modify the experimental setup, due to the associated effort and costs or the impossibility of measuring all the necessary outputs.  
    \item Reduce the dimensions of the model by removing some states or other expressions from the equations. The most common techniques are lumping, time scale separation, sensitivity-based analysis, and singular value decomposition; a recent overview is provided in \cite{snowden2017methods}. These methods are generally not designed to achieve identifiability or observability. 
    Furthermore, most of them are based on previous knowledge of the dynamics, or on the assumption that the evolution of the system can be characterized in different time scales. In addition, they usually yield approximations of the system behavior that are not exact.   
    \item Set certain parameters to a known value found in the literature. This practice makes it possible to reduce the number of parameters to be estimated, and can, it seems, improve the identifiability of the model. However, it must be done with caution, since there is no guarantee that the values to which the parameters are fixed are adequate (bear in mind that if we were reasonably confident about the value of a parameter, we would probably consider it a known constant instead of attempting to estimate it from data).
    \item Reduce the dimension of the model, reparameterizing it to remove the redundant parameters. By eliminating the redundancies, the symmetries in the equations disappear, and the model no longer has the flexibility to compensate for deviations in one parameter with changes in another variable. That is, we obtain a reformulated model that is structurally identifiable and observable.
\end{enumerate}
\end{NewBox}

\begin{enumerate}
    \item Even if we determine which parameter combinations are identifiable, this knowledge does not lead to an automatic reformulation of the model. For example, in the example in Box 1, $p \times s_i$ is an identifiable combination, but this product does not appear in the model and it is therefore not straightforward to replace it. 
    \item Assuming that we manage to reformulate the model so that it becomes identifiable and observable, the transformed variables may lose their mechanistic meaning -- in which case the model cannot provide any insight into them. For the example in Box 1, if the reparameterized model combines insulin ($I$) and $p$ in a new variable $\tilde{I}=I/p$, the new variable may be observable but the model cannot be used to monitor insulin concentration.
\end{enumerate}

To the best of our knowledge, there is currently no method that addresses the two aforementioned challenges satisfactorily. This shortcoming is a major obstacle for the exploitation of mechanistic dynamic models. 
Partial results have been obtained with a number of approaches, including the search for scaling symmetries \cite{MESHKAT2014Identifiable}, 
Gröbner bases \cite{meshkat2014finding}, Taylor Series \cite{evans2000extensions}, similarity transformations \cite{chappell1998procedure}, and Lie Point Symmetries \cite{Merkt2015HigherOrder,Massonis2020Finding}. Although some of the resulting methods have computational implementations, none of them can produce identifiable reparameterizations automatically. The methods based on Lie Point Symmetries are the ones that have come closest to achieving this. Software tools that can be applied to dynamic systems include a Maple implementation \cite{sedoglavic2002probabilistic} that uses a probabilistic seminumerical algorithm with the computation of Hermite-Padé series,
a Python code \cite{Merkt2015HigherOrder} that implements a deterministic method limiting the type of transformations obtained from the variables, and a MATLAB tool \cite{Massonis2020Finding} that extends the previous method to any type of symmetry and computes the transformations using Lie series.
The final three of the above tools can find transformations that break the symmetries of the model equations. However, they introduce an additional parameter that leads to the model variables losing their mechanistic meaning.

Here we present a novel method and an associated software tool (AutoRepar) that together achieve the two aforementioned goals: obtaining an identifiable and observable model reparameterization in a fully automatic way, while preserving the mechanistic meaning of the variables of interest. 
The methodology begins by analyzing a model given by a set of deterministic nonlinear ODEs and an input/output mapping, using the algorithm in \cite{villaverde2019full} to characterize  the structural identifiability of each of its parameters and the observability of each of its state variables. Then, as an intermediate step, it applies the procedure presented in \cite{Massonis2020Finding} to search for the Lie symmetries that cause the lack of identifiability and/or observability. This search goes beyond scaling symmetries, making it possible to find all types of symmetries and to take into account initial conditions and unknown inputs. Our method then builds on this knowledge to compute all possible reparameterizations. Since typically several transformations are possible, the user can select the one that best suits the needs of the application, that is, choosing which parameter(s) can be removed from the model and which ones should be kept. Sometimes it is necessary to apply more than one transformation, in which case the user is consulted at each step. The final result is a reparameterized model that is structurally identifiable and observable. It is provided as a set of equations along with the corresponding changes of variables.

This method yields an exact -- not approximate -- model reformulation without requiring any assumptions about the system behavior. AutoRepar is implemented in MATLAB and integrated in a new version of the widely used STRIKE-GOLDD toolbox \cite{villaverde2019full}.
The method and its computational implementation are the first and, to date, only ones that meet the desired specifications. 
In the next section we illustrate its capabilities by applying it to nonlinear models of different biological disciplines (physiology, cell signaling, pharmacokinetics). We then discuss the main implications of the results, before concluding the paper by providing the methodological details.

\section*{Results}

\begin{figure}[htb]
\centering\includegraphics[width=1\linewidth]{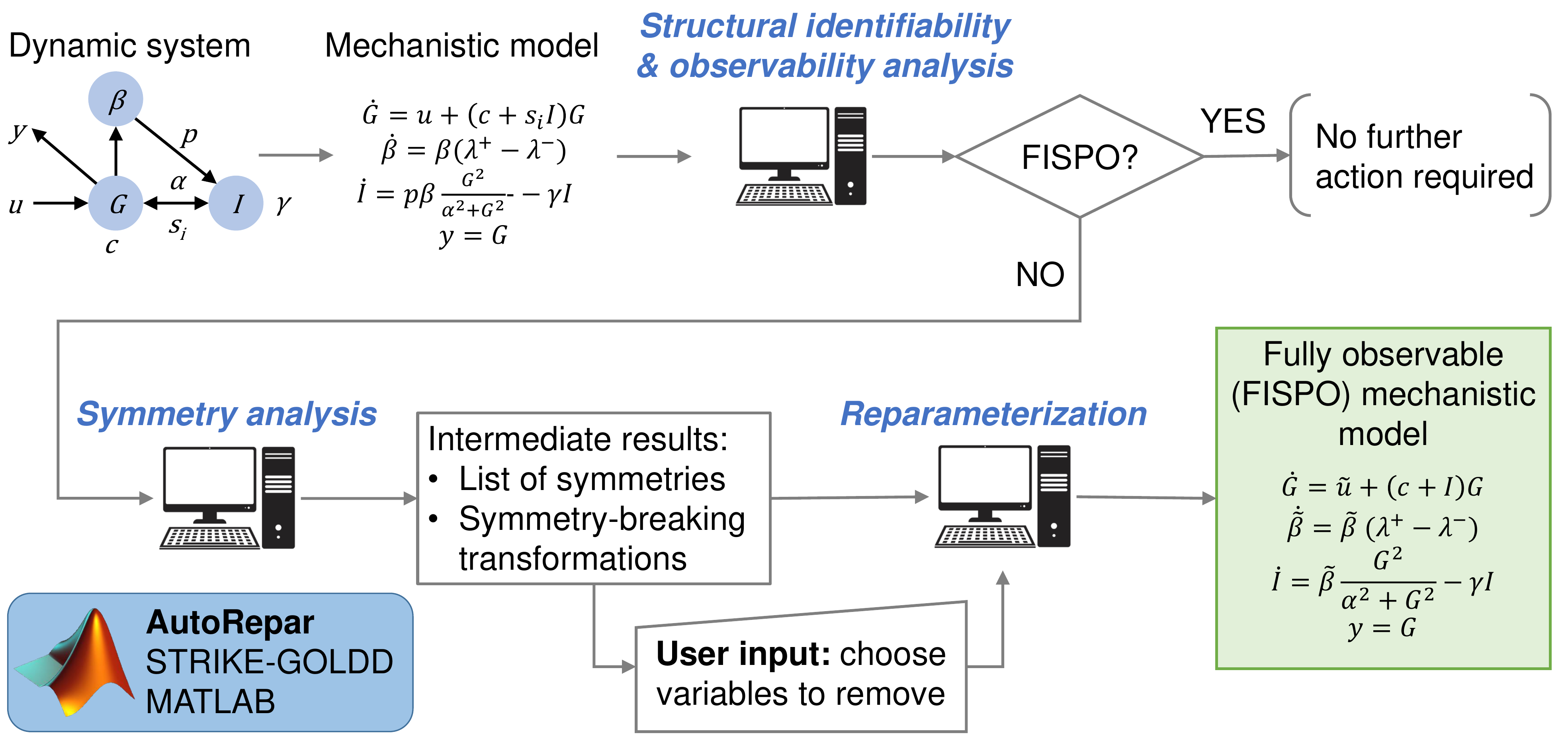}
\caption{\textbf{AutoRepar: automatic reparameterization to obtain identifiable and observable models.} The workflow consists of three main steps. 
The starting point is an ODE model of a dynamical system. The first step is to perform a \textit{structural identifiability and observability analysis}. If the model is structurally identifiable and observable, no further action is required. (Following \cite{villaverde2019full}, we will refer to the property of full input, state, and parameter observability as FISPO.)
If the model is not FISPO, the next step is to perform a \textit{symmetry analysis} to find the existing Lie symmetries and the variable transformations that can be performed to remove them. The third and final step is to perform a model \textit{reparameterization} by applying one or more transformations. This procedure obtains a fully observable version of the original model.
The procedure is implemented in MATLAB and included in the STRIKE-GOLDD toolbox. It is automated and can be performed without user intervention, while allowing for this convenient: when multiple reformulations are possible, the user can choose which parameters or state variables should be removed and which ones should be kept in the model.}
\label{fig_method}
\end{figure}

\paragraph*{Making a model identifiable and observable while preserving its mechanistic meaning.}
We begin with a deliberately simple case study to illustrate the basic functioning of our method. We revisit the first example presented in \cite{vajda1989similarity} by Vajda et al, which is a nonlinear model with an unknown input ($w(t)$), two states ($x_1(t),x_2(t)$), and four unknown parameters ($\theta_1,\theta_2,\theta_3,\theta_4$). The measured output is the first state, $y(t)=x_1(t)$, and the initial conditions are assumed to be zero, $x_1(0)=x_2(0)=0$. The dynamics is given by the following equations:
\begin{align}
    \dot{x}_1 (t) &= w(t)+ \theta_1 x_1 (t)^2+\theta_2 x_1 (t) x_2 (t),\nonumber\\
    \dot{x}_2 (t) &=\theta_3 x_1 (t)^2+\theta_4 x_2 (t) x_1 (t) , \nonumber
\end{align}
We apply our methodology following the workflow shown in Figure \ref{fig_method}. 

First we analyze the SIO of the model, finding that it is non-identifiable and non-observable. Specifically, $\theta_2$ and $\theta_3$ are unidentifiable, and $x_2(t)$ is non-observable.

We then look for the Lie symmetries that cause the lack of identifiability and observability. This analysis finds three infinitesimal generators and their corresponding symmetry-breaking transformations (see Methods for a detailed explanation). 
The first infinitesimal generator involves the unknown input $w$ and the parameter $\theta_2$. By choosing this generator and applying the corresponding transformation we remove the unidentifiable parameter $\theta_2$ from the model, obtaining the following reformulated equations:
\begin{align}
    \dot{x}_1 (t) &= w^{*}(t)+ \theta_1 x_1 (t)^2+ x_1 (t) x_2 (t),\nonumber\\
    \dot{x}_2 (t) &=\theta_3 x_1 (t)^2+\theta_4 x_2 (t) x_1 (t) , \nonumber
\end{align}
where the transformed variable is
$$w^{*}(t)= w(t)+ x_1(t) x_2(t) (\theta_2-1)$$
The reparameterized model shown above is FISPO, i.e. its parameters are structurally identifiable and its unmeasured state and input are observable.
We note that all computations are performed automatically, and the user simply has to choose the preferred transformation from those proposed by the program.

Albeit simple, this case illustrates two interesting features of the method. The first is that \textit{it is capable of finding non-elementary transformations}.
The second is that it manages to \textit{render a variable observable without transforming it}, thus making it possible to infer its value while preserving its mechanistic meaning. This is the case of the state variable $x_2(t)$ and of parameter $\theta_3$, which are observable/identifiable in the final model, unlike in the initial one.

\begin{figure}[H]
\includegraphics[width=1.0\linewidth]{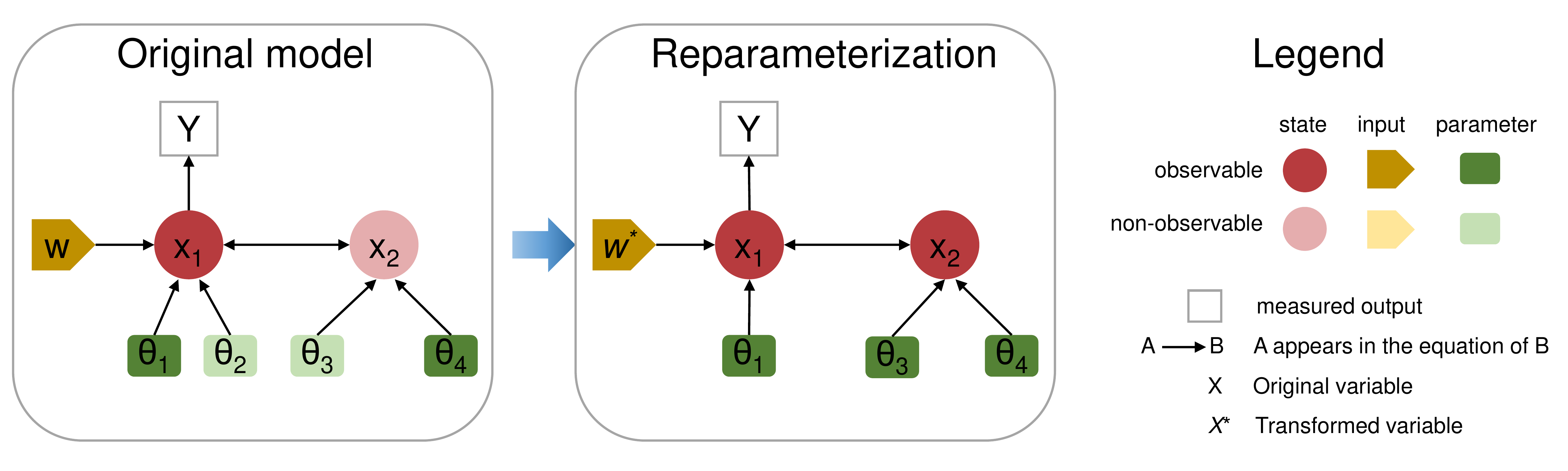}
\caption{The model by Vajda et al \cite{vajda1989similarity}. The image on the left shows the original model (2 states, 4 parameters, 1 output and 1 input function) while the one on the right shows the repaired model (2 states, 3 parameters, 1 output and 1 input function). States are colored in red, inputs in yellow, parameters in green, and outputs in white. The intensity of the colors (darker or lighter) of the first three variables symbolizes whether they are structurally observable or not, respectively. Self-interaction loops (e.g. $x_1$--$x_1$) are not shown.}
\label{fig_vadja}
\end{figure}

Figure \ref{fig_vadja} shows structural diagrams of the original and reformulated models. These diagrams are useful for visualizing which variables are included in each model, along with the relationships among them, and which parts have been simplified. 

It should also be noted that the number of transformations needed to make a model identifiable and observable is given by the difference between the dimension of the augmented state vector and the rank of the observability-identifiability matrix \cite{evans2000extensions,GUNN1997Reparameterisation}. The original formulation of this model has two states, four parameters, one unknown input, and the input derivative, making its dimension equal to eight. (As explained in the Methods section, we assume that the unknown inputs have a finite number of non-zero derivatives. For simplicity we set this number to one; we obtained identical results when setting it to two and to three.)
Since the initial SIO analysis reported that the rank of its observability-identifiability matrix is seven, in this case one transformation was sufficient. Next, we show an example that requires several transformations.

\paragraph*{A model may require several transformations to become identifiable and observable.}
The following example is a pharmacokinetic (PK) model, which describes the time-course of the concentration of a drug after it has been ingested \cite{sedoglavic2002probabilistic}. It consists of four states ($x_1$--$x_4$), nine parameters ($k_1$--$k_7,s_2,s_3$), and an unknown input function ($u$); two scaled states are measured ($y_1=s_2 x_2,y_2=s_3 x_3$):
\begingroup
\allowdisplaybreaks
    \begin{align}
        \dot{x_1}&=u-(k_1+k_2)x_1~,\nonumber \\
        \dot{x_2}&=k_1 x_1-(k_3 + k_6 +k_7)x_2 +k_5 x_4~,\nonumber \\
        \dot{x_3}&= k_2 x_1 +k_3 x_2 -k_4 x_3 ~,\nonumber \\
        \dot{x_4}&=k_6 x_2 -k_5 x_4~,\nonumber  \\
        y_1&=s_2 x_2~,\nonumber \\
        y_2&=s_3 x_3~.\nonumber 
    \end{align}
\endgroup
Note that we have omitted the dependency of $x$ and $u$ on time to simplify the notation.
The initial SIO study classifies only three parameters as identifiable: $k_4$, $k_5$ and $k_6$. Since the augmented state vector has dimension 15 and the rank of the observability-identifiability matrix is 13, it is necessary to apply two transformations to make the model fully identifiable and observable. 

For the first transformation the user can choose from four different infinitesimal generators (full details are provided in the supplementary Information). All the unidentifiable parameters appear in at least one of the symmetries, so any of them can be removed. In this example we choose $k_1$. 

After the first transformation, the second SIO analysis establishes that two parameters have become identifiable, $k_3$ and $k_7$. The subsequent symmetry search returns three possible infinitesimal generators; again, all the remaining unidentifiable parameters are involved in at least one of them. The transformation given by the first generator is useful because it does not require to transform all the states and parameters. This transformation allows us to choose between $k_2$ and $s_3$; we select $s_3$ as the parameter to remove.

\begin{figure}[H]
\includegraphics[width=1.0\linewidth]{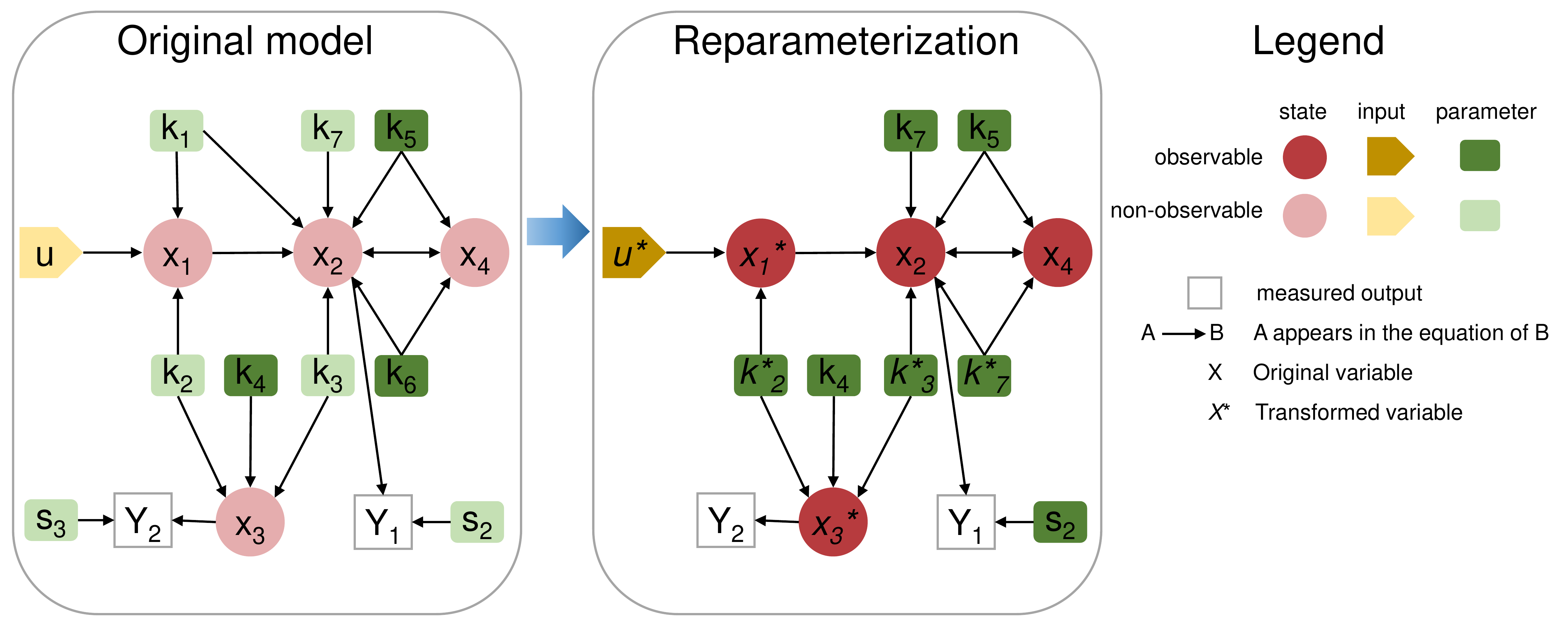}
\caption{Pharmacokinetic model, PK \cite{sedoglavic2002probabilistic}. The image on the left shows the original model (4 states, 9 parameters, 2 outputs and 1 input function) while the one on the right shows the repaired model (4 states, 7 parameters, 2 output and 1 input function). States are colored in red, inputs in yellow, parameters in green, and outputs in white. The intensity of the colors (darker or lighter) of the first three variables symbolizes whether they are structurally observable or not, respectively.  Self-interaction loops (e.g. $x_1$--$x_1$) are not shown.}
\label{fig_pk}
\end{figure}

This procedure yields the following reformulated model:
\begingroup
\allowdisplaybreaks
    \begin{align*}
        \dot{x_1^{*}}&=\widetilde{u}-\widetilde{k_2} x_1^{*}~,\\
        \dot{x_2}&=x_1^{*}-(\widetilde{k_3} + k_6 +\widetilde{k_7})x_2 +k_5 x_4~,\\
        \dot{\widetilde{x_3}}&= \widetilde{k_2} x_1^{*} +\widetilde{k_3} x_2 -k_4 \widetilde{x_3} ~,\\
        \dot{x_4}&=k_6 x_2 -k_5 x_4~, \\
        y_1&=s_2 x_2~,\\
        y_2&=\widetilde{x_3}~.
    \end{align*}
\endgroup
The transformed variables are:
\begingroup
\allowdisplaybreaks
    \begin{align*}
        x_1^{*}&= x_1 k_1,\\
        u^{*}&= x_1 (k_1+k_2)+k_1(u-x_1(k_1+k_2)),\\
        k_2^{*}&= \dfrac{k_2}{k_1},\\
        \widetilde{x_3}&= x_3 s_3,\\
        \widetilde{k_2}&=k_2^{*} s_3= \dfrac{k_2 s_3}{k_1},\\
        \widetilde{k_3}&= k_3 s_3,\\
        \widetilde{k_7}&= k_3 (1-s_3)+k_7,\\
        \widetilde{u}&=\widetilde{k_2} x_1^{*}(s_3-1)+u^{*}=k_2 s_3 x_1 (s_3-1)+x_1 (k_1+k_2)+k_1(u-x_1(k_1+k_2))
    \end{align*}
\endgroup
where we have distinguished with $x_i^{*}$ the changes made in the first transformation and with $\widetilde{x_i}$ those of the second one. Note that the parameter $k_2$ and the input function $u(t)$ have been transformed twice.
The corresponding diagrams are shown in Figure \ref{fig_pk}.

This reformulation eliminates two parameters, $k_1$ and $s_3$, and combines them with $k_2$, $k_3$, $k_7$, $x_1$ and $x_3$. The parameter $s_2$ and the states $x_2$ and $x_4$ become identifiable and observable, respectively, without undergoing any transformation. Thus, we obtain a fully identifiable and observable model without the need to transform all the states and/or parameters. 

\paragraph*{Unknown inputs provide flexibility.}
Let us now examine the $\beta$IG model that was introduced as an example in Box 1. It is composed of three states  ($\beta, I, G$), the latter of which is the measured output (i.e. $y=G$), five parameters ($p,s_i,c,\alpha,\gamma$), and one input, the glucose uptake $u(t)$, which we will start by assuming to be known:
\begingroup
\allowdisplaybreaks
\begin{align*}
    \dot{G}(t) &= u(t)+(c+s_i I(t))G(t), \\
    \dot{\beta}(t) &= \beta (t) (\lambda^{+}-\lambda^{-}),\nonumber \\
    \dot{I} (t)&= p \beta (t) \rho_G -\gamma I(t),\\
    y(t) &= G(t), \\
    \rho_G&=\dfrac{G(t)^2}{\alpha^2+G(t)^2},\\
    \lambda^{+}&=\dfrac{\mu^{+}}{1+\left( \dfrac{8.4}{G(t)}\right)^{1.7}},\\
    \lambda^{-} &= \dfrac{\mu^{-}}{1+\left( \dfrac{G(t)}{4.8}\right)^{8.5}},\\
    \mu^{+}&=\dfrac{0.021}{24 \cdot 60},\\
    \mu^{-}&=\dfrac{0.025}{24 \cdot 60}
\end{align*}
\endgroup 
This model was presented in \cite{Karin2016DynamicalCompensation} and its SIO was analyzed in \cite{villaverde2017dynamical}. The latter study discussed how to make it identifiable and observable by directly measuring additional outputs or parameters. Here we illustrate how to achieve the same goal without requiring additional measurements. 

The initial SIO analysis yields that two parameters are unidentifiable, $p$ and $s_i$, and two states are unobservable, $\beta (t)$ and $I(t)$. Since the rank of the observability-identifiability matrix is 6 and the total number of parameters and states is 8, two transformations are required for eliminating all redundancies.
By removing $s_i$ in the first transformation and $p$ in the second one, we obtain the following reparameterized model:
\begin{align*}
    \dot{G}(t) &= u(t)+(c+ \widetilde{I}(t))G(t), \\
    \dot{\widetilde{\beta}}(t) &= \widetilde{\beta}(t) (\lambda^{+}-\lambda^{-}), \nonumber\\
    \dot{\widetilde{I}} (t)&= \widetilde{\beta}(t) \rho_G -\gamma \widetilde{I}(t), \nonumber\\
    y(t) &= G(t) \nonumber
\end{align*}
where
\begin{align}
    \beta^{*}(t)&= \beta (t) p,\nonumber\\
    \widetilde{\beta}(t)&= \beta (t)^{*} s_i= \beta (t) s_i p,\nonumber\\
    \widetilde{I}(t)&=I(t) s_i  \nonumber
\end{align}
As in the previous example, we write the variables transformed in the first reparameterization with $*$ and those of the second one with $\ \widetilde{}$ .

\begin{figure}[H]
\includegraphics[width=1.0\linewidth]{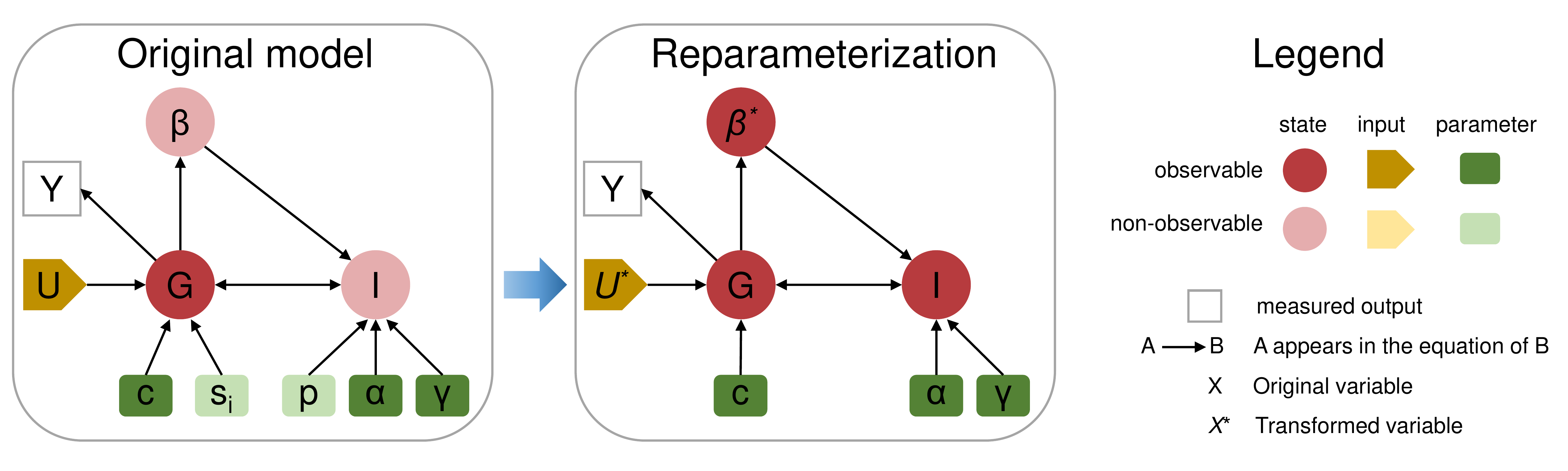}
\caption{$\beta$IG model. The image on the left shows the original model (3 states, 5 parameters, 1 output and 1 input function) while the one on the right shows the repaired model (3 states, 3 parameters, 1 output and 1 input function). States are colored in red, inputs in yellow, parameters in green, and outputs in white. The intensity of the colors (darker or lighter) of the first three variables symbolizes whether they are structurally observable or not, respectively. Self-interaction loops (e.g. $x_1$--$x_1$) are not shown.}
\label{fig_big}
\end{figure}

We note that in this example it has been necessary to transform all the unobservable states and unidentifiable parameters. Thus, while we obtain a fully identifiable and observable model, we cannot use it to infer the variables that were not observable in the original formulation.
However, there is a way of achieving this goal: the idea is to exploit the additional flexibility provided by the assumption that the external input is unknown. Considering unknown input functions has recently been shown to be useful in the context of estimation, in order to account for structural model errors \cite{Newmiwaka2020SEEDS}. Assuming that $u(t)$ is unknown, we obtain a reparameterization that makes the insulin concentration, $I(t)$, observable without undergoing any transformation: 
\begin{align}
    \dot{G}(t) &= u^{*}(t)+(c+ I(t))G(t), \nonumber \\
    \dot{\widetilde{\beta}}(t) &= \widetilde{\beta}(t) (\lambda^{+}-\lambda^{-}), \nonumber\\
    \dot{I} (t)&= \widetilde{\beta}(t) \rho_G -\gamma I(t), \nonumber\\
    y(t) &= G(t) \nonumber
\end{align}
The unknown input assumption allows us to include an additional term in it; reparameterizing the input in this way, along with the $\beta$-cell mass, also eliminates the requirement of transforming the insulin state:
\begin{align*}
    u^{*}(t) & =u(t)+G(t) I(t)(s_i+1),\\
    \widetilde{\beta}(t)&= \beta (t) p
\end{align*}
Figure \ref{fig_big} shows the corresponding diagrams.

\paragraph*{The method can reparameterize models with dozens of states and parameters.}
With this last example we show the applicability of the methodology to models of larger dimension. To this end we consider a model of the transcription factor NF-$\kappa$B signaling with ten states, twenty parameters, four output functions, and one input that we consider unknown \cite{Merkt2015HigherOrder}:
\begingroup
\allowdisplaybreaks
     \begin{align*}
\dot{x}_1 &=k_{11} x_{10}-\left( \dfrac{k_1 u}{1+k_0 u} +k_{1p}\right)x_1~,\\
\dot{x}_2 &=\left( \dfrac{k_1 u}{1+k_0 u}+k_{1p}\right) x_1 -k_2 x_2~,\\
\dot{x}_3 &=k_2 x_2 -k_3 x_3 ~,\\
\dot{x}_4 &= k_2 x_2 -k_4 x_4 ~,\\
\dot{x}_5 &= k_3 \rho_{\text{vol}} x_3 -k_5 x_5 ~, \\
\dot{x}_6&= k_5 x_5 -k_{10} x_9 x_6 ~,\\
\dot{x}_7 &=k_6 x_6 -k_7 x_7 ~,\\
\dot{x}_8 &=k_8 x_7 -k_9 x_8 ~,\\
\dot{x}_9 &= k_9 \rho_{\text{vol}} x_8 -k_{10} x_9 x_6~, \\
\dot{x}_{10} &=k_{10} x_9 x_6 -k_{11}\rho_{\text{vol}} x_{10}~,\\
y_{1} &= s_1 (x_1+x_2+x_3)+I_{0_{\text{cyt}}}~,\\
y_{2} &=s_2 (x_{10}+x_5+x_6) +I_{0_{\text{nuc}}}~,\\
y_{3} &=s_3 (x_2 + x_3)~,\\
y_{4} & = s_4 (x_2+x_4) ~.
     \end{align*}
     \endgroup
     
     This model includes parametric initial conditions:
     \begingroup
\allowdisplaybreaks
     \begin{align*}
x_1 (0)&= x_{1} \nonumber~,\\
x_2 (0) &= \dfrac{k_{1p}x_1 (0)}{k_2}\nonumber~,\\
x_3 (0)& =\dfrac{k_{1p}x_1 (0)}{k_3} \nonumber~,\\
x_4 (0)& =\dfrac{k_{1p}x_1 (0)}{k_4} \nonumber~,\\
x_5 (0)&=\dfrac{k_{1p} \rho_{\text{vol}} x_1 (0)}{k_5}\nonumber~,\\
x_6 (0)&=\dfrac{k_{1p} x_1 (0)}{k_9} \nonumber~, \\
x_7 (0)&= \dfrac{k_6 k_{1p} x_1 (0)}{k_7 k_9}\nonumber~, \\
x_8 (0)&= \dfrac{k_{1p} x_1 (0)}{k_9}\nonumber~, \\
x_9 (0)&= \dfrac{k_9 \rho_{\text{vol}}}{ k_{10}} \nonumber~, \\
x_{10} (0)&= \dfrac{k_{1p} x_1}{k_{11}}  ~.    \label{cond-ini-nfkb}
     \end{align*}
     \endgroup
In this initial formulation only 11 parameters are structurally identifiable ($k_{1p}$, $k_2$, $k_3$, $k_4$, $k_5$, $k_7$, $k_9$, $k_{11}$, $\rho_{\text{vol}}$, $I_{0_{\text{nuc}}}$, and $I_{0_{\text{cyt}}}$). 
The model requires three rounds of transformations to achieve full identifiability and observability. At each step it is possible to choose from several parameters for removal; here we report one possible solution, but others could be selected.

In the first transformation the user can select from $\{ k_0, k_1\}$; we choose to remove $k_0$.
Next, from $\{ k_6, k_8\}$ we remove $k_6$, and finally, from $\{ k_{10}, s_1, s_2, s_3, s_4 \}$ we eliminate $k_{10}$. 

These choices lead to the following model reformulation, the diagram of which is shown in Figure \ref{figure_nfkb}:

\begin{figure}[!t]
\includegraphics[width=1.00\linewidth]{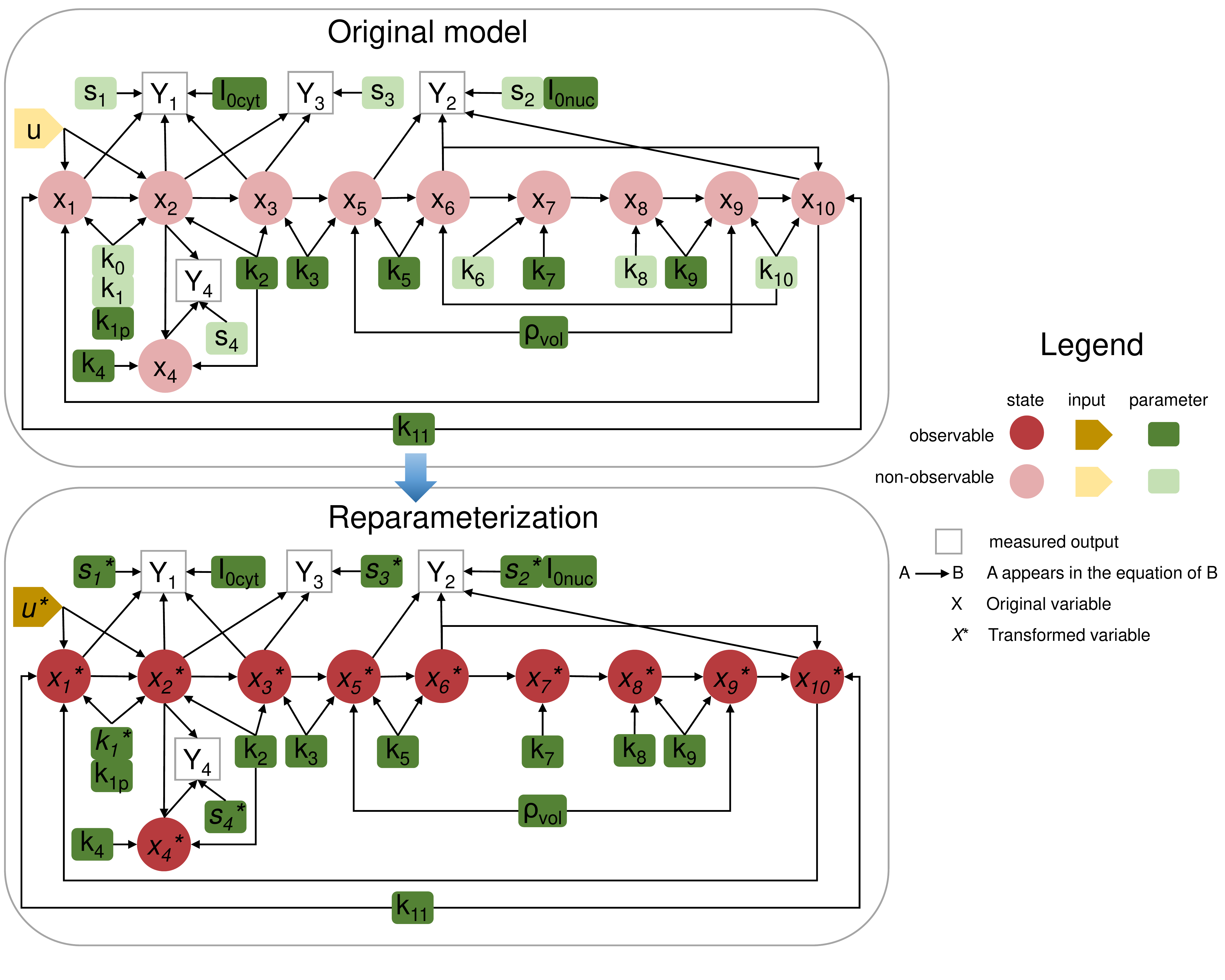}
\caption{NF-$\kappa$B model. To improve readability, only the relationships arising from the dynamic and output equations are shown; the parametrized initial conditions are omitted. The image on the left shows the original model (9 states, 20 parameters, 4 outputs and 1 input function) while the one on the right shows the repaired model (20 states, 17 parameters, 4 output and 1 input function). States are colored in red, inputs in yellow, parameters in green, and outputs in white. The intensity of the colors (darker or lighter) of the first three variables symbolizes whether they are structurally observable or not, respectively. Self-interaction loops (e.g. $x_1$--$x_1$) are not shown.}
\label{figure_nfkb}
\end{figure}

\begingroup
\allowdisplaybreaks
\begin{align*}
\dot{\overline{x_1}} &=k_{11} \overline{x_{10}}-\left( \dfrac{ k_1^{*}u^{*}}{1+ u^{*}} +k_{1p}\right)\overline{x_1}~,\\
\dot{\overline{x_2}} &=\left( \dfrac{k_1^{*}u^{*}}{1+ u^{*}}+k_{1p}\right) \overline{x_1} -k_2 \overline{x_2}~,\\
\dot{\overline{x_3}} &=k_2 \overline{x_2} -k_3 \overline{x_3} ~,\\
\dot{\overline{x_4}} &= k_2 \overline{x_2} -k_4 \overline{x_4} ~,\\
\dot{\overline{x_5}} &= k_3 \rho_{\text{vol}} \overline{x_3} -k_5 \overline{x_5} ~, \\
\dot{\overline{x_6}}&= k_5 \overline{x_5} - \overline{x_9} \overline{x_6} ~,\\
\dot{\overline{x_7}} &=k_6 \overline{x_6} -k_7 \overline{x_7} ~,\\
\dot{\overline{x_8}} &=\widetilde{k_8} \overline{x_7} -k_9 \overline{x_8} ~,\\
\dot{\overline{x_9}} &= k_9 \rho_{\text{vol}} \overline{x_8} - \overline{x_9} \overline{x_6}~, \\
\dot{\overline{x_{10}}} &=\overline{x_9} \overline{x_6} -k_{11}\rho_{\text{vol}} \overline{x_{10}}~,\\
y_{1} &= \overline{s_1} (\overline{x_1}+\overline{x_2}+\overline{x_3})+I_{0_{\text{cyt}}}~,\\
y_{2} &=\overline{s_2} (\overline{x_{10}}+\overline{x_5}+\overline{x_6}) +I_{0_{\text{nuc}}}~,\\
y_{3} &=\overline{s_3} (\overline{x_2} + \overline{x_3})~,\\
y_{4} & = \overline{s_4} (\overline{x_2}+\overline{x_4}) ~.
     \end{align*}
     \endgroup
\noindent where the variables transformed in the first ($x^{*}$), second ($\widetilde{x}$), and third ($\overline{x}$) reparameterizations are, respectively:
\begin{align*}
    u^{*}&= u k_0, ~k_1^{*}=\dfrac{k_1}{k_0};\\
    \widetilde{x_7}&=\dfrac{x_7}{k_6},~\widetilde{k_8}=k_8 k_6;\\
    \overline{x_i}&= x_i k_{10}~\text{for}~i=1,...,6,8,9,10,~\overline{x_7}=\widetilde{x_7}k_{10}=\dfrac{x_7 k_{10}}{k_6},~\overline{s_j}= \dfrac{s_j}{k_{10}}~\text{for}~j=1,...,4;
\end{align*}

\section*{Discussion}

A lack of structural identifiability and observability (SIO) can seriously hamper the usefulness of a dynamic model, and a great deal of research over recent decades has focused on addressing this issue, leading to computational tools capable of analyzing SIO ever more efficiently. Yet the answer to a related question has remained elusive: how to fix the identifiability and observability issues detected by these tools systematically and efficiently? 

With this in mind, we have presented here a methodology that automatically provides a fully identifiable and observable reparameterization of an otherwise non-identifiable and/or non-observable model.
Our method, called AutoRepar, achieves this goal by eliminating those symmetries in the model equations that cause the lack of SIO. Thus, a prerequisite for its application is that the SIO of the model has been analyzed and any existing symmetries have been found. To account for these requirements we have paired the new method with two of our previous developments for (i) analyzing SIO and (ii) finding the Lie symmetries that cause the lack of SIO. The combination of these three techniques yields an integrated procedure that completes the ``model curation'' process, as far as SIO is concerned. First, the SIO of a model is analyzed; second, if the analysis reports lack of SIO, the Lie symmetries that cause the problem are sought; third, the symmetries are automatically repaired. 

AutoRepar is applicable to nonlinear ODE models with rational expressions, which are a very general class of models used widely in many areas of the life sciences. AutoRepar makes these models fully identifiable and observable through a reparameterization that has a number of remarkable features. First, it keeps the output functions invariant, i.e. it does not require measuring additional states. Second, it is an exact reformulation, not an approximation, i.e. the dynamics of the final model is identical to that of the original one. Third, while the reparameterization typically removes a subset of the parameters, and transforms a number of model variables (parameters and/or states), it usually leaves another subset of variables intact. Thus, these variables that are not transformed preserve their mechanistic meaning. 
Fourth, the method can be applied to models with unknown inputs. 
Indeed, we have found that unknown inputs provide flexibility to the reparameterization, and in some cases the consideration of an input as an unknown function can make it possible to keep some variables from being transformed. 
Fifth, it can consider parameterized initial conditions.
Sixth, it can repair symmetries of different types, not only the simplest ones such as scaling.
Seventh, it can repair models that need several transformations, each of which may be a non-elementary transformation.

We have implemented the AutoRepar methodology in a new version of our open source MATLAB toolbox, STRIKE-GOLDD, which already included routines for analyzing SIO and finding Lie symmetries. The resulting tool provides an integrated platform that analyzes the structural identifiability and observability of a model, finds the causes that prevent them, and repairs them.
The tool is fully automated: after entering the model equations, no further input is necessary from the user. However, the user can enter input if desired, in order to select which variables (from a given subset) can be removed or transformed, and which ones must be preserved. This feature is crucial since it allows for specifying which predictions about parameters or states are to be obtained from the model.

We have illustrated the methodological aspects discussed above using four dynamic models of pharmacokinetics, physiological, and signaling processes. These case studies have also demonstrated the applicability of the computational implementation, showing that the tool can reparameterize models with dozens of variables. 

\section*{Methods}

\paragraph*{Notation and definitions of structural identifiability and observability.}
We consider models of deterministic ordinary differential equations (ODEs) of the form:
\begin{numcases}{\mathcal{M}=}\label{din}
	\dot{x}(t)=f\left(x(t),\theta,u(t),w(t)\right)\\\label{out}
	y(t)=g\left(x(t),\theta,u(t),w(t)\right)
\end{numcases}
where $f$ and $g$ are analytical functions of the states $x(t)\in\mathbb{R}^{n_x},$ known inputs $u(t)\in\mathbb{R}^{n_u}$, unknown constant parameters $\theta\in\mathbb{R}^{n_\theta}$, and unknown inputs $w(t)\in\mathbb{R}^{n_w}$. The output $y(t)\in\mathbb{R}^{n_y}$ represents the measurable functions of model variables.

A parameter $\theta_i$ of model $\mathcal{M}$ is \textit{structurally locally identifiable} (s.l.i.) if for almost any parameter vector $\theta^*\in\mathbb{R}^{n_\theta}$ there is a neighborhood ${\mathcal N}(\theta^*)$ in which the following relationship holds \cite{distefano2015dynamic}:
\begin{equation}\label{eq:sli}
	\hat{\theta} \in {\mathcal N}(\theta^*) \text{ and } y(t,\hat{\theta}) = y(t,\theta^*) \Rightarrow \hat{\theta_i} = \theta_i^*
\end{equation}
If the above implication is not fulfilled, $\theta_i$ is \textit{structurally unidentifiable} (s.u.).
If all model parameters are s.l.i. the model is s.l.i. If there is at least one s.u. parameter, the model is s.u..   
Similarly, a state $x_i(\tau)$ is \textit{observable} if it can be distinguished from any other states in a neighborhood from observations of the model output $y(t)$ and input $u(t)$ in the interval $t_0 \leq \tau \leq t \leq t_f$, for a finite $t_f$. Otherwise, $x_i(\tau)$ is \textit{unobservable}. A model is called observable if all its states are observable. Finally, $\mathcal{M}$ is \textit{input observable} if it is possible to infer its unknown input $w(t)$, which is said to be observable or reconstructible. 

\paragraph*{Full Input-State-Parameter Observability (FISPO).}
Identifiability and input observability can be considered particular instances of a general observability property. To reflect this view, the acronym FISPO, which stands for Full Input-State-Parameter Observability (or Observable when used as an adjective) has been introduced \cite{villaverde2019full}. Its definition is as follows.
Consider a model $\mathcal{M}$ given by (\ref{din}--\ref{out}). 
Let $z(t)=\begin{pmatrix}x(t)&\theta &w(t)\end{pmatrix}$ be the vector of its states, parameters, and unknown inputs. $\mathcal{M}$ has the FISPO property if, for every $t_0$ in the time interval $\left[t_0,t_f\right]\subset .$, every model unknown $z_i(\tau)$ can be inferred from $y(t)$ and $u(t)$. Thus, $\mathcal{M}$ is FISPO if, for every $z(t_0)$, for almost any vector $z^\ast(t_0),$ there is a neighborhood $\mathcal{N}\left(z^\ast\left(t_0\right)\right)$ such that, for all $\hat{z}(t_0)\in\mathcal{N}\left(z^\ast\left(t_0\right)\right),$ the following condition holds:
\begin{align*}
&y\left(t,\hat{z}(t_0)\right)=y\left(t,z^\ast\left(t_0\right)\right)\Rightarrow \hat{z}_i\left(t_0\right)=z_i^\ast\left(t_0\right),\quad 1\leq i \leq n_x+n_\theta+n_w.
\end{align*}

\paragraph*{Assessing the property FISPO with a differential geometry approach.}
Structural identifiability can be studied in conjunction with observability by considering the unknown parameters $\theta$ as state variables with zero dynamics, which leads to an augmented state vector ${\widetilde x}=(x^T,\theta^T)^T$ \cite{tunali1987new}. The reconstructibility of unknown inputs $w(t)$ can also be seen as a particular case of observability, although in this case their derivatives may be nonzero.
To this end we augment the state vector further with $w$ as additional states, as well as their derivatives up to some non-negative integer $l$:
\begin{align}\label{x_aug}
&{\widetilde x(t)}=\begin{pmatrix}
x(t)^T&
\theta^T&
w(t)^T&
\dots&
w(t)^{\left(l\right)^T}\end{pmatrix}^T,
\end{align}
The $l-$augmented dynamics is:
\begin{align*}
&\dot {\widetilde x}(t)=f^l\left({\widetilde x}(t),u(t)\right)=\begin{pmatrix}
f\left(x(t),\theta,u(t),w(t)\right)^T&
0_{1\times n_\theta}&
\dot{w}(t)^T&
\dots&
w^{\left(l+1\right)}(t)^T
\end{pmatrix}^T,
\end{align*}
leading to the $l-$augmented system:
\begin{align}\label{aug_sys}
\mathcal{M}^l
&\begin{cases}
&\dot {\widetilde x}(t)=f^l({\widetilde x}(t),u(t))\\
&y(t)=g({\widetilde x}(t),u(t))
\end{cases}
\end{align}

From now on we omit the dependency of the time-varying variables on time, to simplify the notation.
The property FISPO is assessed by calculating the rank of a generalized observability matrix constructed with ``extended'' Lie derivatives, which are given by \cite{villaverde2019full}:

\begin{align*}
&L_{\tilde{f}}^h\left(\tilde{x},u\right)=\frac{\partial h}{\partial \tilde{x}}\left(\tilde{x},u\right) \tilde{f}\left(\tilde{x},u\right)+\frac{\partial g}{\partial u}\left(\tilde{x},u\right) \dot{u}.
\end{align*}

The zero-order derivative is $L^{0}_{\tilde{f}}g=g,$ and the $i-$order extended Lie derivatives can be recursively calculated as:

\begin{align*}
&L^{i}_{\tilde{f}}h\left(\tilde{x},u\right)=\frac{\partial L^{i-1}_{\tilde{f}}h}{\partial \tilde{x}}\left(\tilde{x},u\right) \tilde{f}\left(\tilde{x},u\right)+\sum_{j=0}^{i-1}\frac{\partial L^{i-1}_{\tilde{f}}g}{\partial u^{\left.j\right)}}\left(\tilde{x},u\right) u^{\left.j+1\right)},\quad i\geq 1.
\end{align*}

The observability-identifiability matrix of $\mathcal{M}$ (\ref{aug_sys}) is:
\begin{align}\label{matr_obs}
&\mathcal{O}_I\left(\tilde{x},u\right)=\frac{\partial}{\partial \tilde{x}}
\begin{pmatrix}
L^0_{\tilde{f}}h\left(\tilde{x},u\right)^T&L_{\tilde{f}}h\left(\tilde{x},u\right)^T&L^2_{\tilde{f}}h\left(\tilde{x},u\right)^T&\dots& L^{n_{\tilde{x}}-1}_{\tilde{f}}h\left(\tilde{x},u\right)^T
\end{pmatrix}^T,
\end{align}

We can now state the following Observability-Identifiability Condition (OIC): a model $\mathcal{M}$ defined by (\ref{aug_sys}) is FISPO around a (possibly generic) point in the augmented state space $\tilde{x}_0$ if the rank of its observability-identifiability matrix (\ref{matr_obs}) is: rank$\left(\mathcal{O}_I\left(\tilde{x}_0,u\right)\right)=n_{\tilde{x}} = n_x+n_\theta+n_w$.

If the observability-identifiability matrix is rank-deficient the model is overparameterized, that is, there are parameters that must be eliminated to achieve the property FISPO. The following theorem gives the number of parameters that must be eliminated from the model:

\begin{theorem}[Existence of reparameterization \cite{GUNN1997Reparameterisation}] Let $Y (\theta_1,..., \theta_p,t)$ be a function whose Taylor series expansion gives rise to the coefficients $h_1 (\theta),...,h_p(\theta)$, where $h_i : A \subset \mathcal{R}^p \Longrightarrow \mathcal{R}$ for $i=1,..,p$. Define $H=(h_1,...,h_p)$. If the Jacobian Matrix $DG(\theta)$ with respect to $\theta$ has rank $q(<p)$ for all $\theta$ in a neighborhood of $\theta^{0} \in A$, then the function $Y$ may be locally reparameterized in terms of a set of $q$ of the Taylor series coefficients,
\begin{equation}
    \{ \phi_1 =h_{J_1},..., \phi_q =h_{J_q}\}
\end{equation}
i.e.
\begin{equation}
    H(\theta_1,..., \theta_p)= \widetilde{H} (\phi_1,...,\phi_q)
\end{equation}
The reparameterized system is locally identifiable.
\end{theorem}

The identifiability of each parameter is assessed by comparing the rank of the observability-identifiability matrix before and after eliminating the column corresponding to that parameter. If rank remains constant the parameter is unidentifiable, while if it decreases it is identifiable. The same procedure is applied to assess the observability of each state.

\paragraph*{Finding the roots of non-observability: Lie symmetries.}
The lack of structural identifiability and/or observability means that the model output does not vary if some parameters or states are modified in a certain way.
The presence of Lie symmetries in the model equations amounts to the existence of similarity transformations that allow the parameters and states to be transformed while leaving the output invariant \cite{yates2009structural}. Thus, the presence of transformations of this type in the model can be used to characterize the relationships that cause the lack of structural identifiability and/or observability. In what follows we provide a brief introduction to the study of Lie symmetries; more details can be found elsewhere, for example in \cite{bluman2008symmetry}.

A one-parameter Lie group of transformations is a morphism that maps a solution of the differential equations onto itself in terms of state variables, $x^{*}=X(x;\varepsilon)$. 
Its expansion in some neighborhood of $\varepsilon=0$ is:
\begin{equation}\label{infi_lie}
    x^{*}=x+\varepsilon \left( \dfrac{\partial X(x;\varepsilon)}{\partial \varepsilon} \vert_{\varepsilon=0}\right)+\dfrac{1}{2}\varepsilon^2 \left( \dfrac{\partial^2 X(x;\varepsilon)}{\partial \varepsilon^2} \vert_{\varepsilon=0}\right)+...=x+\varepsilon \left( \dfrac{\partial X(x;\varepsilon)}{\partial \varepsilon} \vert_{\varepsilon=0}\right)+O(\varepsilon^2)~,
\end{equation}
The \textit{infinitesimal} of the Lie group of transformations (\ref{infi_lie}) is:
\begin{equation}
\label{eta_expr}
  \eta (x) =\dfrac{\partial X(x; \varepsilon)}{\partial \varepsilon} \vert_{\varepsilon=0} ~
\end{equation}
and $x+ \varepsilon \eta (x)$ is the \textit{infinitesimal transformation}. 

The \textit{infinitesimal generator} is the differential operator: 
\begin{equation}
\label{generador}
    X=X(x)=\eta (x) \cdot \bigtriangledown=\sum_{i=1}^{n}\eta_{i}(x)\dfrac{\partial}{\partial x_{i}} ~,
\end{equation}
where $\bigtriangledown$ defines the gradient 
\begin{equation}
    \bigtriangledown = \left(\dfrac{\partial}{\partial x_{1}}, \dfrac{\partial}{\partial x_{2}}, ..., \dfrac{\partial}{\partial x_{n}} \right) ~.
\end{equation}

The one-parameter Lie group of transformations (\ref{infi_lie}) is equivalent to:
\begin{equation}
\label{lie_series}
    x^{*}=\exp[\varepsilon X]x=x+\varepsilon X x+\dfrac{1}{2}\varepsilon^2 X^2 x+...=
\end{equation}
$$=\left( 1+ \varepsilon X + \dfrac{1}{2} \varepsilon^2 X^2+...\right)x=\sum_{k=0}^{\infty }\dfrac{\varepsilon^{k}}{k!}X^{k}x ~,$$
where $X$ is given by (\ref{generador}) and $X^{k}=XX^{k-1}, ~ k=1,2...$ with $X^{0}x=x$.

To find the Lie symmetries of a dynamic model $\mathcal{M}$ (\ref{aug_sys}), we augment its state vector as in (\ref{x_aug}), with dimension $n_{\tilde{x}}=n_\theta+n_x+n_w$:
\begin{equation*}
    \begin{split}
         \dot{x}_{i}(t)&=f_{i}(x(t),u(t)), \hspace{0.2cm} i=1,...,n_x\\
        x_{i}(t)&=\theta, \hspace{0.2cm} i=m+1,...,n_x+n_\theta\\
        x_{i}(t)&=w_{i}(t), \hspace{0.2cm} i=n_x+n_\theta+1,...,n_{\tilde{x}}
    \end{split}
\end{equation*}
 Then, the following criterion is applied:
 
 \begin{theorem}
\label{teorema1}
\cite{bluman2008symmetry}
The system of ordinary differential equations admits a one-parameter Lie group of transformations defined by the infinitesimal generator \text{(\ref{generador})} if and only if:
\begin{equation}
    X'\cdot (\dot{x}_{k}-f_{k}(x))=0, \hspace{0.2cm} k=1,...,n_x
\end{equation}
\begin{equation}
    X \cdot (y_{l}-g_{l}(x))=0, \hspace{0.2cm} l=1,...,n
\end{equation}
\end{theorem}
Then, an admitted Lie symmetry is a continuous group of transformations $X$ such that the observed data is unchanged:

$$g(x^{*}(t),u^{*}(t))= g(x(t),u(t)) $$

Thus, the output map should not be modified. 

The first step to find Lie symmetries is the creation of (\ref{eta_expr}). Different expressions will be considered depending on the complexity of the combinations of states, parameters and/or unknown input functions. These expressions are \textit{Ansatz polynomials} where the coefficients are the abovementioned combinations and its variables are unknown constants to determine. Then, the polynomials are incorporated in Theorem (\ref{teorema1}) leading to an overdetermined linear system of equations with numeric entries, whose kernel is the infinitesimal generators. The last step is to compute the one-parameter Lie group of transformations with Lie series (\ref{lie_series}).  For a more detailed explanation, see \cite{Massonis2020Finding}.

\paragraph*{Obtaining identifiable and observable reparameterizations with mechanistic interpretations.} \label{Choose_Right}
To make a model observable and identifiable the number of its variables must be equal to the rank of its observability-identifiability matrix. When a transformation removes a non-identifiable parameter, the rank of the matrix remains the same while the number of variables decreases, shortening the distance between both values.
The application of the one-parameter Lie groups of transformations described in the preceding paragraph can render a model structurally identifiable and observable, i.e. FISPO. However, it also modifies the parameter(s) and/or state(s) involved in the transformation(s), introducing an artificial parameter $\varepsilon$ in their equations that makes them lose their mechanistic meaning \cite{Massonis2020Finding}. To obtain a reformulation of the model that preserves its mechanistic character, AutoRepar expresses $\varepsilon$ in terms of other parameters, that is, it normalizes one of the transformed parameters to the unit value and removes $\varepsilon$ from the expression.

We illustrate how this procedure works with the example from Vajda et al \cite{vajda1989similarity}. In the analysis of this model, shown in the Results section, we removed the parameter $\theta_2$ by introducing it into the unknown input function $w$.
The one-parameter transformations are:
$$ w^{*}=x_1 x_2 \varepsilon+w, ~\theta_2^{*}=-\varepsilon+\theta_2$$
Since we want to remove $\theta_2$, we find an expression for $\varepsilon$ for which the parameter $\theta_2^{*}=1$. This expression is given by:
$$\varepsilon=\theta_2 -1 ~, $$
which provides the following reparameterization:
$$w^{*}= w+ x_1 x_2 (\theta_2-1), ~\theta_2^{*}=1 ~ ;$$

Another example used in the Results, the pharmacokinetic (PK) model, requires several transformations to become FISPO. In the first one, the one-parameter Lie groups are:
\begin{align*}
    x_1^{*}&=x_1 \exp{(\varepsilon)}, \\
    u^{*} & = x_1 (k_1+k_2)+\exp{(\varepsilon)}(u-x_1 (k_1+k_2)), \\
    k_1^{*} &= k_1 \exp{(-\varepsilon)},\\
    k_2^{*} &= k_2 \exp{(-\varepsilon)}
\end{align*}
In this step it is possible to remove either $k_1$ or $k_2$. The easiest way to remove the artificial parameter is not by isolating $\varepsilon$ but $\exp{(\varepsilon)}$. For example, if we want to remove $k_1$ we could do:
$$ k_1^{*} = k_1 \exp{(-\varepsilon)}=1 \longrightarrow \exp{(\varepsilon)}= k_1 $$

After reparameterizing the model in this way we perform a second symmetry search, obtaining the following one-parameter Lie-groups of transformations:
\begin{align*}
    \widetilde{x_3}&= x_3 \exp{(\varepsilon)},\\
    \widetilde{k_2}&=k_2^{*} \exp{(\varepsilon)},\\
    \widetilde{k_3}&= k_3 \exp{(\varepsilon)},\\
    \widetilde{k_7}&=k_3 (1-\exp{(\varepsilon)})+k_7,\\
    \widetilde{s_3}&=s_3 \exp{(-\varepsilon)},\\
    \widetilde{u}&= \widetilde{k_2} x_1^{*}(\exp{(\varepsilon)}-1)+u^{*}
\end{align*}
We remove the parameter $s_3$ by isolating $\exp{(\varepsilon)}$ from the above expression: $$\widetilde{s_3}=s_3 \exp{(-\varepsilon)}=1 \longrightarrow \exp{(\varepsilon)}=s_3$$ 
The reparameterized model is:
\begin{align*}
    \widetilde{x_3}&= x_3 s_3,\\
    \widetilde{k_2}&=k_2^{*} s_3,\\
    \widetilde{k_3}&= k_3 s_3,\\
    \widetilde{k_7}&=k_3 (1-s_3)+k_7,\\
    \widetilde{s_3}&=\dfrac{s_3}{s_3}=1,\\
    \widetilde{u}&= \widetilde{k_2} x_1^{*}(s_3-1)+u^{*}
\end{align*}
This model has undergone two reparameterizations to become FISPO. The second one involves two parameters that already appeared in the first one, $k_2$ and $u$. We can express these two variables as functions of the original ones by recursively replacing the transformations in reverse order:
\begin{align*}
\widetilde{k_2}&=k_2^{*} s_3 = \dfrac{k_2}{k_1}s_3,\\
\widetilde{u}&=\widetilde{k_2} x_1^{*}(s_3-1)+u^{*}=\dfrac{k_2}{k_1} s_3 x_1 k_1 (s_3-1)+ x_1 (k_1+k_2)+k_1(u-x_1 (k_1+k_2))=\\
&= k_2 s_3 x_1 (s_3-1)+x_1(k_1+k_2)+k_1 (u-x_1 (k_1+k_2))
\end{align*}

\paragraph*{Computational implementation: a new version of the STRIKE-GOLDD toolbox.}
We have implemented AutoRepar in a new release (v3.0) of the STRIKE-GOLDD toolbox, which is available at GitHub:
\href{https://github.com/afvillaverde/strike-goldd/releases/tag/v3.0}{https://github.com/afvillaverde/strike-goldd/releases/tag/v3.0}.
STRIKE-GOLDD is an open source Matlab toolbox for analyzing nonlinear ODE models. It determines the structural identifiability of their parameters and the observability of their states and unknown inputs, and it finds the Lie symmetries that cause the lack of these properties. The new version is also capable of providing fully observable reparameterizations. All the analyses reported in this paper have been carried out using STRIKE-GOLDD v3.0.

\paragraph*{Supplementary information.}
Supplementary material for this article is available at the file
AutoRepar\_supplementary\_information.pdf, which can be found in the doc folder of STRIKE-GOLDD v3.0 (\href{https://github.com/afvillaverde/strike-goldd/releases/tag/v3.0}{https://github.com/afvillaverde/strike-goldd/releases/tag/v3.0}).

\bibliographystyle{ieeetr}
\bibliography{biblio_ident.bib,biblio_models.bib,biblio_other.bib}
\noindent \textbf{Acknowledgements:} 
This work was inspired by discussions at the Banff International Research Station workshop on ``Model Theory of Differential Equations, Algebraic Geometry, and their Applications to Modeling'', held in 2020. The authors especially thank Gleb Pogudin for helpful conversations.
\textbf{Funding:} This research has received funding from the Spanish Ministry of Science, Innovation and Universities and the European Union FEDER under project grant SYNBIOCONTROL (DPI2017-82896-C2-2-R) and the CSIC intramural project grant MOEBIUS (PIE 202070E062). The funding bodies played no role in the design of the study, the collection and analysis of data, or in the writing of the manuscript.
\textbf{Author Contributions:} A.F.V. designed the study. G.M. developed the code and performed the computational analyses. All authors discussed the results and wrote the manuscript.
\textbf{Competing Interests:} The authors declare that they have no competing financial interests.
\textbf{Data and materials availability:} All data needed to evaluate the conclusions in the paper are present in the paper and/or the supplementary information. Models, codes, and instructions are available from \href{https://github.com/afvillaverde/strike-goldd/releases/tag/v3.0}{https://github.com/afvillaverde/strike-goldd/releases/tag/v3.0}.

\end{document}